\documentclass[a4paper]{article}

\usepackage[utf8]{inputenc}
\usepackage[top=10mm]{geometry}
\usepackage{booktabs}
\usepackage[flushleft]{threeparttable}
\usepackage{authblk}
\usepackage{bm}
\usepackage{amsmath}
\usepackage{amssymb}
\usepackage{graphicx}
\usepackage[square,sort,numbers]{natbib}
\allowdisplaybreaks[4]

\title{Peripheral nucleon-nucleon scattering at next-to-next-to-leading order in SU(3) heavy baryon chiral perturbation theory}

\author[1]{Bo-Lin Huang \thanks{blhuang@pku.edu.cn}}
\author[1]{Jian-Bo Cheng \thanks{jbcheng@pku.edu.cn}}
\author[1]{Shi-Lin Zhu \thanks{zhusl@pku.edu.cn}}

\affil[1]{\normalsize School of Physics and Center of High Energy Physics, Peking University, Beijing 100871, China}

\date{\today}

\begin{document}
\maketitle

\begin{abstract}
We calculate the complete $T$ matrices of the elastic nucleon-nucleon scattering up to third order in SU(3) heavy baryon chiral perturbation theory. The phase shifts with orbital angular momentum $L\geq 2$ and the mixing angles with $J\geq 2$ are evaluated by using low-energy constants that were extracted from the meson-baryon analysis. It turns out that our prediction is consistent with the empirical phase shifts and mixing angles data, and also the results from SU(2) heavy baryon chiral perturbation theory. The errors from the low-energy constants are analyzed in detail. Our calculation provides reliable evidence that the nucleon-nucleon interaction calculated in SU(3) heavy baryon chiral perturbation theory leads to reasonable predictions.
\begin{description}
\item[Keywords:]
Chiral perturbation theory, nucleon-nucleon scattering, SU(3) symmetry
\end{description}
\end{abstract}

\section{Introduction}
The nucleon-nucleon interaction is one of the fundamental problems in nuclear physics and nuclear astrophysics. As the fundamental theory of strong interaction, quantum chromodynamics (QCD) becomes nonperturbative at low energies; thus, it is very difficult to use perturbative methods to derive the nucleon-nucleon interaction. In order to solve this problem, an effective field theory (EFT) was proposed by Weinberg in a seminal paper \cite{wein1979}. The EFT is formulated in terms of the most general Lagrangian consistent with the general symmetry principles, particularly the chiral symmetry of QCD. At low energy, the degrees of freedom are hadrons, i.e. pions, kaons, eta-mesons and baryons, rather than quarks and gluons, while heavy mesons and baryon resonances are integrated out. The corresponding field theoretical formalism is called chiral perturbation theory (ChPT) \cite{sche2012}. ChPT is an efficient framework to calculate in a model-independent way, e.g., the amplitudes of the nucleon-nucleon scattering below the chiral symmetry breaking scale $\Lambda_\chi\sim 1$ GeV. Nevertheless, a power-counting problem in baryon ChPT occurs because of the nonvanishing baryon mass $M_0$ in the chiral limit. The heavy baryon chiral perturbation theory (HB$\chi$PT) has been proposed and developed to solve the power-counting problem \cite{jenk1991,bern1992}. The chiral expressions in HB$\chi$PT proceed simultaneously in terms of $p/\Lambda_\chi$ and $p/M_0$, where $p$ denotes the meson momentum (or mass) or the small residual momentum of a baryon in a low-energy process. The infrared regularization of the covariant baryon ChPT \cite{bech1999}  and the extended-on-mass-shell scheme \cite{gege1999,fuch2003} for baryon ChPT are two popular relativistic approaches, and have led to substantial progress in many aspects as documented in Refs.~\cite{schi2007,geng2008,mart2010,ren2012,alar2012,alar2013}. However, the expressions from the loop diagrams become rather complicated\cite{chen2013,yao2016,lu2019}. On the other hand, HB$\chi$PT is a well-established and versatile tool for the study of low-energy hadronic process, e.g., the nuclear force in SU(2) HB$\chi$PT.

Since the derivation of the nuclear force from chiral effective field theory was proposed by Weinberg in the 1990s \cite{wein1990,wein1991}, many researchers became interested in the field \cite{ordo1992,cele1992,ordo1994,ball1998,kais1998,
epel1998,kais2001,ente2002,ente2015,wu2018,kais2019}. Even the full relativistic amplitudes have been derived in chiral effective field theory \cite{higa2003,higa2004,higa2008,ren2018,xiao2020,ren2021}. However, these amplitudes for nuclear force only involved pions and nucleons in the SU(2) case. For processes involving kaons or hyperons, e.g., the nucleon-hyperon interaction, the hyperon-hyperon interaction, etc., the situation becomes more complicated because of the consequences of three-flavor chiral dynamics. The hyperon-nucleon interaction has been calculated up to next-to-leading order in Refs.~\cite{haid2013,haid2020}. Nevertheless, good convergence cannot be achieved at this order in ChPT, and the higher-order contributions should be considered, but this will involve many more low-energy constants. In previous works \cite{huan2015,huan2017,huan20201,huan20202}, we have investigated the meson-baryon scattering in SU(3) HB$\chi$PT by fitting to partial-wave phase shifts of the pion-nucleon and kaon-nucleon scattering. In particular, the separated low-energy constants $b_i$ from the second-order Lagrangian were obtained by fitting the phase shifts of pion-nucleon scattering at fourth order in SU(3) HB$\chi$PT. Meanwhile, there is a large amount of data for nucleon-nucleon scattering. As an attempt to study baryon-baryon interaction at higher order, we study the nucleon-nucleon scattering at next-to-next-to-leading order in SU(3) HB$\chi$PT in this paper. The peripheral partial waves ($L\geq 2$ and $J\geq 2$) are considered. As for the lower partial waves, they are dominated by the dynamics at the short-range part of the nuclear force. Moreover, the contact terms which consist of four-nucleon fields and no meson fields are used to parametrize the short-distance dynamics. Thus, the lower partial waves are almost the same as in the SU(2) case and are not involved in this paper. Our study provides reliable evidence that the nucleon-nucleon interaction calculated in SU(3) HB$\chi$PT leads to reasonable predictions.

The present paper is organized as follows. In Sec.~\ref{lagrangian}, we summarize the Lagrangians involved in the evaluation of the third-order contributions. In Sec.~\ref{tmatrices}, we present some explicit expressions for the $T$ matrices of the elastic nucleon-nucleon scattering at order $p^3$. In Sec.~\ref{phaseshifts}, we outline how to calculate the phase shifts and mixing angles from the $T$ matrices. Section~\ref{results} contains the presentation and discussion of our results. A short summary is given in the last section.

\section{Chiral Lagrangian}
\label{lagrangian}
In order to calculate the peripheral nucleon-nucleon scattering amplitudes up to order $\mathcal{O}(p^3)$ in heavy baryon SU(3) chiral perturbation theory, the corresponding effective Lagrangian can be written as
\begin{align}
\label{lag}
\mathcal{L}_{\text{eff}}=\mathcal{L}^{(2)}_{\phi \phi}+\mathcal{L}^{(1)}_{\phi B}+\mathcal{L}^{(2)}_{\phi B},
\end{align}
where the superscript indicates the number of derivatives or small external momenta or meson mass. The traceless Hermitian $3\times 3$ matrices $\phi$ and $B$ include the pseudoscalar Goldstone boson fields ($\pi$, $K$, $\bar{K}$, $\eta$) and
the octet-baryon fields ($N$, $\Lambda$, $\Sigma$, $\Xi$), respectively. The lowest-order SU(3) chiral Lagrangians for meson-meson interaction take the form \cite{gass1985}
\begin{align}
\label{lagphiphi}
\mathcal{L}^{(2)}_{\phi\phi}=\frac{f^2}{4}\text{tr}(u_\mu u^\mu + \chi_{+}),
\end{align}
where $f$ is the pseudoscalar decay constant in the chiral limit. The axial vector quantity $u^\mu=i\{\xi^{\dagger},\partial^\mu\xi\}$ contains odd number meson fields. The quantity $\chi_{+}=\xi^{\dagger}\chi\xi^{\dagger}+\xi\chi\xi$ with $\chi=\text{diag}(m_\pi^2,m_\pi^2,2m_K^2-m_\pi^2)$ introduces explicit chiral symmetry breaking terms. The SU(3) matrix $U=\xi^2=\text{exp}(i\phi/f)$ collects the pseudoscalar Goldstone boson fields. Note that, the so-called sigma parametrization was chosen in SU(2) HB$\chi$PT \cite{mojz1998,fett1998}. The lowest-order chiral meson-baryon heavy baryon Lagrangian \cite{bora1997} is
\begin{align}
\label{lagphiB1}
 \mathcal{L}_{\phi B}^{(1)}=\text{tr}(i\overline{B}[v\cdot D,B])+D\,\text{tr}(\overline{B}S_{\mu}\{u^{\mu},B\})+F\,\text{tr}(\overline{B}S_{\mu}[u^{\mu},B]),
\end{align}
where $D_{\mu}$ denotes the chiral covariant derivative
\begin{align}
\label{dmuB}
[D_{\mu},B]=\partial_{\mu}B+[\Gamma_{\mu},B],
\end{align}
and $S_{\mu}$ is the covariant spin operator
\begin{align}
\label{spino}
S_\mu=\frac{i}{2}\gamma_5 \sigma_{\mu\nu}v^\nu,\quad S\cdot v=0,
\end{align}
\begin{align}
\label{spinor}
\{S_\mu,S_\nu\}=\frac{1}{2}(v_\mu v_\nu-g_{\mu\nu}),\quad [S_\mu,S_\nu]=i\epsilon_{\mu\nu\sigma\rho}v^\sigma S^\rho,
\end{align}
where $\epsilon_{\mu\nu\sigma\rho}$ is the complete antisymmetric tensor with $\epsilon_{0123}=1.$ The chiral connection $\Gamma^\mu=[\xi^{\dagger},\partial^\mu\xi]/2$ contains even number meson fields. The axial vector coupling constants $D$ and $F$ can be determined from semileptonic hyperon decays \cite{bora1999}.

Beyond the leading order, the complete heavy baryon Lagrangian splits up into two parts,
\begin{align}
\label{lagphiBi}
\mathcal{L}_{\phi B}^{(i)}=\mathcal{L}_{\phi B}^{(i,\text{rc})}+\mathcal{L}_{\phi B}^{(i,\text{ct})}\quad (i\ge 2),
\end{align}
where $\mathcal{L}_{\phi B}^{(i,\text{rc})}$ denotes $1/M_0$ expansions with fixed coefficients and stems from the original relativistic Lagrangian. Here, $M_0$ stands for the (average) octet mass in the chiral limit. The remaining heavy baryon Lagrangian $\mathcal{L}_{\phi B}^{(i,\text{ct})}$ is proportional to the low-energy constants.

The heavy baryon Lagrangian $\mathcal{L}_{\phi B}^{(2,\text{ct})}$ can be obtained from the relativistic effective meson-baryon chiral Lagrangian \cite{olle2006,frin2006}
\begin{align}
\label{lagphi2ct}
\mathcal{L}_{\phi B}^{(2,\text{ct})}=&\,\,b_{D}\,\text{tr}(\overline{B}\{\chi_{+},B\})+b_{F}\,\text{tr}(\overline{B}[\chi_{+},B])
+b_{0}\,\text{tr}(\overline{B}B)\text{tr}(\chi_{+})+b_{1}\,\text{tr}(\overline{B}\{u^{\mu}u_{\mu},B\})\nonumber\\
&+b_{2}\,\text{tr}(\overline{B}[u^{\mu}u_{\mu},B])+b_{3}\,\text{tr}(\overline{B}B)\text{tr}(u^{\mu}u_{\mu})
+b_{4}\,\text{tr}(\overline{B}u^{\mu})\text{tr}(u_{\mu}B)+b_{5}\,\text{tr}(\overline{B}\{v\cdot u\,\,v\cdot u,B\})\nonumber\\
&+b_{6}\,\text{tr}(\overline{B}[v\cdot u\,\,v\cdot u,B])+b_{7}\,\text{tr}(\overline{B}B)\text{tr}(v\cdot u\,\,v\cdot u)
+b_{8}\,\text{tr}(\overline{B}v\cdot u )\text{tr}( v\cdot u B)\nonumber\\
&+b_{9}\,\text{tr}(\overline{B}\{[u^{\mu},u^{\nu}],[S_{\mu},S_{\nu}]B\})
+b_{10}\,\text{tr}(\overline{B}[[u^{\mu},u^{\nu}],[S_{\mu},S_{\nu}]B])\nonumber\\
&+b_{11}\,\text{tr}(\overline{B}u^{\mu})\text{tr}(u^{\nu}[S_{\mu},S_{\nu}]B),
\end{align}
The first three terms of $\mathcal{L}_{\phi B}^{(2,\text{ct})}$ are proportional to the LECs $b_{D,F,0}$ and result in explicit symmetry breaking. All LECs $b_i$ have dimension $\text{mass}^{-1}$.

The $\mathcal{L}_{\phi B}^{(2,\text{rc})}$ reads
\begin{align}
\label{lagphiB2rc}
\mathcal{L}_{\phi B}^{(2,\text{rc})}=\,\,&\frac{D^{2}-3F^{2}}{24M_{0}}\text{tr}(\overline{B}[v\cdot u,[v\cdot u,B]])
-\frac{D^{2}}{12M_{0}}\text{tr}(\overline{B}B)\text{tr}(v\cdot u\,\, v\cdot u)\nonumber\\
&-\frac{DF}{4M_{0}}\text{tr}(\overline{B}[v\cdot u,\{v\cdot u,B\}])
-\frac{1}{2M_{0}}\text{tr}(\overline{B}[D_{\mu},[D^{\mu},B]])\nonumber\\
&+\frac{1}{2M_{0}}\text{tr}(\overline{B}[v\cdot D,[v\cdot D,B]])
-\frac{iD}{2M_{0}}\text{tr}(\overline{B}S_{\mu}[D^{\mu},\{v\cdot u,B\}])\nonumber\\
&-\frac{iF}{2M_{0}}\text{tr}(\overline{B}S_{\mu}[D^{\mu},[v\cdot u,B]])
-\frac{iF}{2M_{0}}\text{tr}(\overline{B}S_{\mu}[v\cdot u,[D^{\mu},B]])\nonumber\\
&-\frac{iD}{2M_{0}}\text{tr}(\overline{B}S_{\mu}\{v\cdot u,[D^{\mu},B]\}).
\end{align}
Since we explicitly work out the various $1/M_0$ expansions, the last three terms of $\mathcal{L}_{\phi B}^{(2,\text{rc})}$ are not absorbed in the corresponding LECs $b_i$.

\section{$T$ matrix for the nucleon-nucleon scattering}
\label{tmatrices}
\begin{figure}[t]
\centering
\includegraphics[height=12cm,width=13cm]{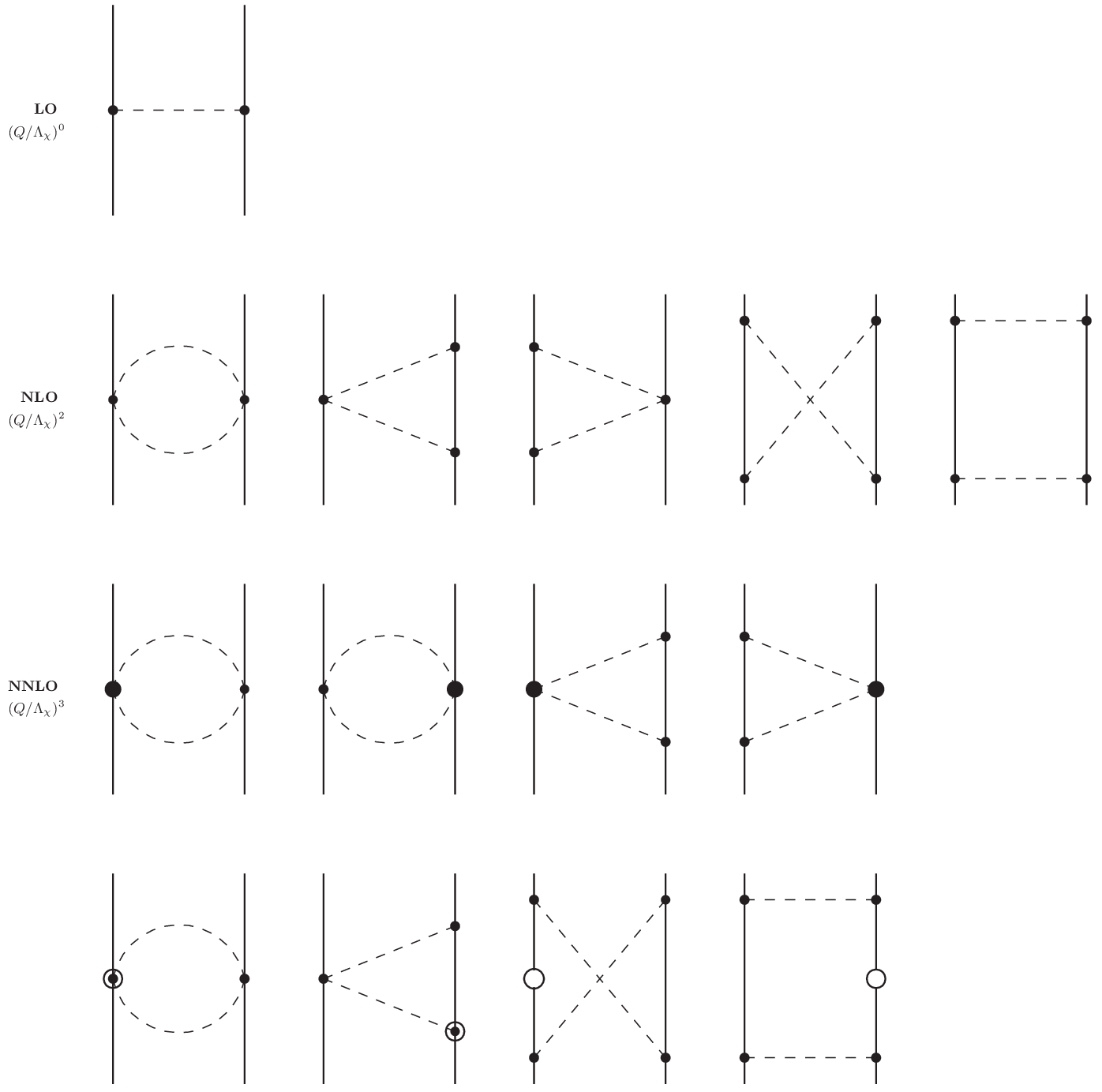}
\caption{\label{fig:nnloopfeynman}LO, NLO, and NNLO contributions to the $NN$ interaction. Solid lines represent baryons and dashed lines mesons. Small dots denote vertices from the leading-order Lagrangian $\mathcal{L}_{\phi B}^{(1)}$. Large solid dots are vertices proportional to the LECs $b_i$ from the second-order Lagrangian $\mathcal{L}_{\phi B}^{(2)}$. Symbols with an open circle are relativistic $1/M_0$ corrections which are also part of $\mathcal{L}_{\phi B}^{(2)}$. Only a few representative examples of $1/M_0$ corrections are shown. }
\end{figure}

We are considering only elastic nucleon-nucleon scattering processes $N(\vec{p}\,)+N(-\vec{p}\,)\rightarrow N(\vec{p}^{\,'})+N(-\vec{p}^{\,'})$ in the center-of-mass system (c.m.s.), where $\pm\vec{p}$ and $\pm\vec{p}^{\,'}$ are the initial and final nucleon momenta in the c.m.s., respectively. In the center-of-mass frame, the elastic on-shell momentum-space nucleon-nucleon $T$ matrix has the general form
\begin{align}
\label{tmatrice}
T(\vec{p}\,',\vec{p}\,)=&V_C+\bm{\tau_1}\cdot\bm{\tau_2}W_C+[V_S+\bm{\tau_1}\cdot\bm{\tau_2}W_S]\vec{\sigma}_1\cdot\vec{\sigma}_2
+[V_T+\bm{\tau_1}\cdot\bm{\tau_2}W_T]\vec{\sigma}_1\cdot\vec{q}\,\vec{\sigma}_2\cdot\vec{q}\nonumber\\
&+[V_{LS}+\bm{\tau_1}\cdot\bm{\tau_2}W_{LS}][-i\vec{S}\cdot(\vec{q}\times \vec{k})]+[V_{\sigma L}+\bm{\tau_1}\cdot\bm{\tau_2}W_{\sigma L}]\vec{\sigma}_1\cdot(\vec{q}\times\vec{k})\,\vec{\sigma}_2\cdot(\vec{q}\times\vec{k}),
\end{align}
where $\vec{q}=\vec{p}\,'-\vec{p}$ is the momentum transfer, $\vec{k}=(\vec{p}\,'+\vec{p}\,)/2$ is the average momentum, and $\vec{S}=(\vec{\sigma}_1+\vec{\sigma}_2)/2$ is the total spin, with $\vec{\sigma}_1$, $\vec{\sigma}_2$ and $\bm{\tau_1}$, $\bm{\tau_2}$ being the spin and isospin operators, of nucleons 1 and 2, respectively. For the on-shell scattering, the isoscalar $V_\alpha$ and isovector $W_\alpha (\alpha=C,S,T,LS,\sigma L)$ can be expressed as functions of $q=|\vec{q}\,|$ and $p=|\vec{p}\,'|=|\vec{p}\,|$ only. The momentum transfer $q=2p\,\text{sin}\frac{\theta}{2}=p\sqrt{2(1-z)}$ in the center-of-mass system with $z=\text{cos}(\theta)$, the cosine of the angle $\theta$ between $\vec{p}$ and $\vec{p}^{\,'}$. Our notation and conventions follow Ref.~\cite{mach2011}, which is commonly used in nuclear physics.

In order to get a more compact representation of the amplitudes, the four combinations of the axial vector coupling constants $D$ and $F$ are introduced,
\begin{align}
\label{DFcom}
g_A=& D+F, \nonumber\\
g_\eta=& D-3F,\nonumber\\
g_\alpha=& (D+3F)^2+9(D-F)^2,\nonumber\\
g_\beta=& (D+3F)^2-3(D-F)^2,
\end{align}
where $g_A$ is the usual axial vector coupling constants in the SU(2) case, and $g_\eta$ is related to the $\eta$-meson coupling constants.

In the following, we present the analytical results for the nucleon-nucleon $T$ matrix up to third order. These amplitudes are divided into two parts where the contributions are from the contact terms and the meson exchange. The local contact terms are not considered in this paper, since they do not contribute to the phase shifts with $L\geq 2$ and mixing angles with $J\geq 2$. We will now derive the meson-exchange contributions to the nucleon-nucleon interaction order by order. The leading-order (LO, $\nu=0$) amplitudes resulting from the first row one-meson-exchange diagram in Fig.~\ref{fig:nnloopfeynman} read
\begin{align}
\label{VWLO}
V_T^{(\text{LO})}=-\frac{g_\eta^2}{12f_\eta^2(q^2+m_\eta^2)},\quad W_T^{(\text{LO})}=-\frac{g_A^2}{4f_\pi^2(q^2+m_\pi^2)},
\end{align}
with the physical values of the meson-nucleon coupling constant, meson mass and decay constants. Our results are consistent with the amplitudes calculated in the SU(2) case \cite{kais19971}, except that the contributions from $\eta$-meson exchange are included. The contributions from vertex and propagator corrections to one-meson exchange are not considered since they only contribute to mass and coupling constant renormalization. As usual, the counting power $\nu$ is given by
\begin{align}
\label{power}
\nu=2l+\sum_{i}\Big(d_i+\frac{n_i}{2}-2\Big),
\end{align}
where $l$ denotes the number of loops in the diagram, $d_i$ the number of derivatives or meson-mass insertions and $n_i$ the number of nucleon fields involved in vertex $i$. The sum runs over all vertices $i$ in the diagram under consideration.

At next-to-leading-order (NLO, $\nu=2$), the leading two-meson-exchange diagrams appear; see the second row of Fig.~\ref{fig:nnloopfeynman}. All diagrams can be calculated in a straightforward manner by using the heavy baryon formalism except for the planar box diagram, since the diagram involves the iterated one-meson-exchange contribution. Note that we include only the noniterative part of this diagram which is obtained by subtracting the iterated one-meson-exchange contribution. After a tedious calculation, the amplitudes from irreducible two-meson exchange at this order read
\begin{align}
\label{VCNLO}
V_C^{\text{(NLO)}}=&-\frac{1}{165888\pi^2f_K^4(4m_K^2+q^2)}\Big\{(4m_K^2+q^2)[48(162+9g_\alpha+2g_\alpha^2)m_K^2+(90+g_\alpha)(18
\nonumber\\
&+5g_\alpha)q^2]+6[64(2g_\alpha^2-18g_\alpha-81)m_K^2+16(7g_\alpha^2-63g_\alpha-162)m_K^2q^2+(23g_\alpha^2-180g_\alpha\nonumber\\&-324)q^4]L(m_K,m_K)+6(4m_K^2+q^2)[18(5g_\alpha^2-36g_\alpha-108)m_K^2+(23g_\alpha^2-180g_\alpha\nonumber\\
&-324)q^2]\text{ln}\frac{m_K}{\lambda}\Big\},
\end{align}

\begin{align}
\label{WCNLO}
W_C^{\text{(NLO)}}=&\frac{1}{2304\pi^2f_\pi^4(4m_\pi^2+q^2)}\Big\{(4m_\pi^2+q^2)[24(1+g_A^2+4g_A^4)m_\pi^2+(5+g_A^2)(1+5g_A^2)q^2]\nonumber\\&+6[16(1+4g_A^2-8g_A^4)m_\pi^4+8(1+7g_A^2-14g_A^4)m_\pi^2q^2+(1+10g_A^2\nonumber\\
&-23g_A^4)q^4]L(m_\pi,m_\pi)+6(4m_\pi^2+q^2)[6(-1-6g_A^2+15g_A^4)m_\pi^2+(-1-10g_A^2\nonumber\\
&+23g_A^4)q^2]\text{ln}\frac{m_\pi}{\lambda}\Big\}-\frac{1}{165888\pi^2f_K^4(4m_K^2+q^2)}\Big\{(4m_K^2+q^2)[48(18+3g_\beta+2g_\beta^2)m_K^2
\nonumber\\
&+(30+g_\beta)(6+5g_\beta)q^2]+6[64(2g_\beta^2-6g_\beta-9)m_K^2+16(7g_\beta^2-21g_\beta-18)m_K^2q^2\nonumber\\&+(23g_\beta^2-60g_\beta-36)q^4]L(m_K,m_K)+6(4m_K^2+q^2)[18(5g_\beta^2-12g_\beta-12)m_K^2\nonumber\\
&+(23g_\beta^2-60g_\beta-36)q^2]\text{ln}\frac{m_K}{\lambda}\Big\},
\end{align}

\begin{align}
\label{VTVSNLO}
V_T^{\text{(NLO)}}=-\frac{1}{q^2}V_S=&\frac{3g_A^4}{128\pi^2f_\pi^4}\Big\{1-2L(m_\pi,m_\pi)-2\text{ln}\frac{m_\pi}{\lambda}\Big\}+\frac{g_\alpha^2}{4608\pi^2f_K^4}\Big\{1-2L(m_K,m_K)\nonumber\\
&-2\text{ln}\frac{m_K}{\lambda}\Big\}+\frac{g_\eta^4}{1152\pi^2 f_\eta^4}\Big\{1-2L(m_\eta,m_\eta)-2\text{ln}\frac{m_\eta}{\lambda}\Big\},
\end{align}

\begin{align}
\label{WTWSNLO}
W_T^{\text{(NLO)}}=-\frac{1}{q^2}W_S=&\frac{g_\beta^2}{4608\pi^2f_K^4}\Big\{1-2L(m_K,m_K)-2\text{ln}\frac{m_K}{\lambda}\Big\}+\frac{g_A^2g_\eta^2}{384\pi^2f_\eta^2f_\pi^2q^2}\Big\{2q^2\nonumber\\
&+2(m_\eta^2-m_\pi^2)\text{ln}\frac{m_\eta}{m_\pi}-2q^2[L(m_\eta,m_\pi)+L(m_\pi,m_\eta)]-4q^2\text{ln}\frac{\sqrt{m_\eta m_\pi}}{\lambda}\Big\},
\end{align}
where we have defined the functions
\begin{align}
\label{Lfunction}
L(m_1,m_2)=&\frac{w(m_1,m_2)}{2q}\text{ln}\frac{[q w(m_1,m_2)+q^2]^2-(m_1^2-m_2^2)^2}{4m_1m_2 q^2},\nonumber\\
w(m_1,m_2)=&\frac{1}{q}\sqrt{[q^2+(m_1+m_2)^2][q^2+(m_1-m_2)^2]}.
\end{align}

At next-to-next-to-leading-order (NNLO, $\nu=3$), the resulting amplitudes are either proportional to one of the low-energy constants $b_i$ or they contain a factor $1/M_0$. The relativistic $1/M_0$ corrections arise from vertices and baryon propagators. In Fig.~\ref{fig:nnloopfeynman}, the diagrams with one vertex proportional to $b_i$ (large solid dot) are shown in the third row, and the four representative diagrams with a $1/M_0$ correction (symbols with an open circle) are shown in the fourth row. The number of $1/M_0$ correction diagrams is large and not all of them are shown. Note that all football diagrams vanish at this order. Again, all but the planar box diagram can be calculated in a straightforward manner using the heavy baryon formalism. Following the method of finishing first the $l_0$ integral and then expanding in $1/M_0$ from Ref.~\cite{kais19971}, the $1/M_0$ corrections from the planar box diagram can be calculated quickly and correctly. Then, the full third-order amplitudes read
\begin{align}
\label{VCNNLO}
V_C^{\text{(NNLO)}}=&\frac{3g_A^4}{256\pi M_0 f_\pi^4(4m_\pi^2+q^2)}\Big\{5m_\pi^5+13m_\pi^3q^2+3m_\pi q^4+3(8m_\pi^4q^2+6m_\pi^2q^4\nonumber\\
&+q^6)A(m_\pi,m_\pi)\Big\}+\frac{3g_A^2}{16\pi f_\pi^4}\Big\{-2(2b_0-2b_1-2b_2-4b_3+b_D+b_F)m_\pi^3+(b_1+b_2\nonumber\\
&+2b_3)m_\pi q^2-(2m_\pi^2+q^2)[2(2b_0-b_1-b_2-2b_3+b_D+b_F)m_\pi^2-(b_1+b_2\nonumber\\
&+2b_3)q^2]A(m_\pi,m_\pi)\Big\}+\frac{g_\alpha}{4608\pi M_0 f_K^4(4m_K^2+q^2)}\Big\{(-288+26g_\alpha)m_K^5+18(-8\nonumber\\&+g_\alpha)m_K^3q^2+3(-6+g_\alpha)m_K q^4+[2(-18+g_\alpha)m_K^2+3(-6+g_\alpha)q^2](8m_K^4\nonumber\\
&+6m_K^2q^2+q^4)A(m_K,m_K)\Big\}+\frac{1}{96\pi f_K^4}\Big\{2[-2(2b_0-3b_1+b_2-4b_3-b_4+b_D)g_\alpha\nonumber\\
&+(-b_D+b_F)g_\beta]m_K^3+(3b_1-b_2+4b_3+b_4)g_\alpha m_K q^2-(2m_K^2+q^2)[2(4b_0-3b_1\nonumber\\
&+b_2-4b_3-b_4+2b_D)g_\alpha m_K^2+2(b_D-b_F)g_\beta m_K^2-(3b_1-b_2+4b_3\nonumber\\
&+b_4)g_\alpha q^2]A(m_K,m_K)\Big\}+\frac{g_\eta^4}{2304\pi M_0 f_\eta^4(4m_\eta^2+q^2)}\Big\{5m_\eta^5+13m_\eta^3q^2+3m_\eta q^4\nonumber\\
&+3(8m_\eta^4q^2+6m_\eta^2q^4+q^6)A(m_\eta,m_\eta)\Big\}+\frac{g_\eta^2}{144\pi f_\eta^4}\Big\{-16(b_0+b_D-b_F)m_K^2m_\eta\nonumber\\
&+2(2b_0+3b_D-5b_F)m_\pi^2m_\eta+(5b_1-3b_2+6b_3)(4m_\eta^2+q^2)m_\eta+(2m_\eta^2\nonumber\\
&+q^2)[-16(b_0+b_D-b_F)m_K^2+2(2b_0+3b_D-5b_F)m_\pi^2+(5b_1-3b_2+6b_3)(2m_\eta^2\nonumber\\
&+q^2)]A(m_\eta,m_\eta)\Big\},
\end{align}

\begin{align}
\label{WCNNLO}
W_C^{\text{(NNLO)}}=&\frac{g_A^2}{128\pi M_0 f_\pi^4(4m_\pi^2+q^2)}\Big\{(-32+47g_A^2)m_\pi^5+(-16+23g_A^2)m_\pi^3q^2+(-2+3g_A^2)m_\pi q^4\nonumber\\
&+[4(g_A^2-1)m_\pi^2+(3g_A^2-2)q^2](8m_\pi^4+6m_\pi^2q^2+q^4)A(m_\pi,m_\pi)\Big\}\nonumber\\
&+\frac{g_\beta}{4608\pi M_0 f_K^4(4m_K^2+q^2)}\Big\{(-96+26g_\beta)m_K^5+6(-8+3g_\beta)m_K^3q^2+3(-2\nonumber\\&+g_\beta)m_K q^4+[2(-6+g_\beta)m_K^2+3(-2+g_\beta)q^2](8m_K^4+6m_K^2q^2+q^4)A(m_K,m_K)\Big\}\nonumber\\
&+\frac{g_\beta}{96\pi f_K^4}\Big\{2(2b_1+2b_2+2b_4-b_D-b_F)m_K^3+(b_1+b_2+b_4)m_K q^2+(2m_K^2\nonumber\\
&+q^2)[2(b_1+b_2+b_4-b_D-b_F)m_K^2+(b_1+b_2+b_4)q^2]A(m_K,m_K)\Big\}\nonumber\\
&+\frac{g_A^2g_\eta^2}{768\pi M_0 f_\pi^2f_\eta^2[(m_\eta+m_\pi)^2+q^2]}\Big\{(m_\eta+m_\pi)[m_\eta^4+m_\eta^3m_\pi+m_\eta^2m_\pi^2+m_\eta m_\pi^3\nonumber\\
&+m_\pi^4+(4m_\eta^2+5m_\eta m_\pi+4m_\pi^2)q^2+3q^4]+3q^2(m_\eta^2+m_\pi^2+q^2)[(m_\eta+m_\pi)^2\nonumber\\
&+q^2][A(m_\eta,m_\pi)+A(m_\pi,m_\eta)]\Big\}-\frac{g_A g_\eta}{48\pi f_\pi^2f_\eta^2}\Big\{(m_\eta+m_\pi)[-2(b_D+b_F)m_\pi^2\nonumber\\&+(b_1+b_2)(3m_\eta^2-2m_\eta m_\pi+3m_\pi^2+q^2)]+(m_\eta^2+m_\pi^2+q^2)[-2(b_D+b_F)m_\pi^2\nonumber\\
&+(b_1+b_2)(m_\eta^2+m_\pi^2+q^2)][A(m_\eta,m_\pi)+A(m_\pi,m_\eta)]\Big\},
\end{align}

\begin{align}
\label{VTVSNNLO}
V_T^{\text{(NNLO)}}=-\frac{1}{q^2}V_S=&\frac{9g_A^4}{512\pi M_0 f_\pi^4}\Big\{m_\pi+(2m_\pi^2+q^2)A(m_\pi,m_\pi)\Big\}-\frac{g_A^2}{16\pi f_\pi^4}\Big\{(b_9+b_{10})[m_\pi\nonumber\\&+(4m_\pi^2+q^2)A(m_\pi,m_\pi)]\Big\}+\frac{g_\alpha}{9216\pi M_0 f_K^4}\Big\{3(-6+g_\alpha)m_K+[8(-9\nonumber\\
&+g_\alpha)m_K^2+3(-6+g_\alpha)q^2]A(m_K,m_K)\Big\}+\frac{g_\alpha}{384\pi f_K^4}\Big\{(2b_9-6b_{10}\nonumber\\
&-b_{11})[m_K+(4m_K^2+q^2)A(m_K,m_K)]\Big\}+\frac{g_\eta^4}{1536\pi M_0 f_\eta^4}\Big\{m_\eta+(2m_\eta^2\nonumber\\
&+q^2)A(m_\eta,m_\eta)\Big\},
\end{align}

\begin{align}
\label{WTWSNNLO}
W_T^{\text{(NNLO)}}=-\frac{1}{q^2}W_S=&\frac{g_A^2}{256\pi M_0 f_\pi^4}\Big\{(-2+3g_A^2)m_\pi+[2(-4+5g_A^2)m_\pi^2+(-2\nonumber\\
&+3g_A^2)q^2]A(m_\pi,m_\pi)\Big\}+\frac{g_\beta}{9216\pi M_0 f_K^4}\Big\{3(-2+g_\beta)m_K+[8(-3+g_\beta)m_K^2\nonumber\\
&+3(-2+g_\beta)q^2]A(m_K,m_K)\Big\}-\frac{g_\beta}{384\pi f_K^4}\Big\{(2b_9+2b_{10}+b_{11})[m_K\nonumber\\
&+(4m_K^2+q^2)A(m_K,m_K)]\Big\}+\frac{g_A^2g_\eta^2}{512\pi M_0 f_\pi^2f_\eta^2}\Big\{m_\eta+m_\pi+(m_\eta^2+m_\pi^2\nonumber\\
&+q^2)[A(m_\eta,m_\pi)+A(m_\pi,m_\eta)]\Big\},
\end{align}

\begin{align}
\label{VLSNNLO}
V_{LS}^{\text{(NNLO)}}=&\frac{3g_A^4}{32\pi M_0 f_\pi^4}\Big\{m_\pi+(2m_\pi^2+q^2)A(m_\pi,m_\pi)\Big\}+\frac{g_\alpha}{2304\pi M_0 f_K^4}\Big\{(18+g_\alpha)m_K+[72m_K^2\nonumber\\
&+(18+g_\alpha)q^2]A(m_K,m_K)\Big\}+\frac{g_\eta^4}{288\pi M_0 f_\eta^4}\Big\{m_\eta+(2m_\eta^2+q^2)A(m_\eta,m_\eta)\Big\},
\end{align}

\begin{align}
\label{WLSNNLO}
W_{LS}^{\text{(NNLO)}}=&\frac{g_A^2(1-g_A^2)}{32\pi M_0 f_\pi^4}\Big\{m_\pi+(4m_\pi^2+q^2)A(m_\pi,m_\pi)\Big\}+\frac{g_\beta}{2304\pi M_0 f_K^4}\Big\{(6+g_\beta)m_K+[24m_K^2\nonumber\\
&+(6+g_\beta)q^2]A(m_K,m_K)\Big\}+\frac{g_A^2g_\eta^2}{96\pi M_0 f_\eta^2f_\pi^2}\Big\{m_\eta+m_\pi+(m_\eta^2+m_\pi^2+q^2)[A(m_\eta,m_\pi)\nonumber\\
&+A(m_\pi,m_\eta)]\Big\},
\end{align}
where
\begin{align}
\label{Afunction}
A(m_1,m_2)=\frac{1}{2q}\text{arctan}\frac{q}{m_1+m_2}.
\end{align}

At last, we calculate the reducible two-meson-exchange contribution which is generated from iterated one-meson exchange. The iterated one-meson exchange is the only contribution to the nucleon-nucleon $T$ matrix with a nonvanishing imaginary part and restores unitarity at this order. In fact, the planar box diagram includes the iterated one-meson exchange, which has the following integral representation:
\begin{align}
\label{itint}
V_{\text{it}}(m_1,m_2)=&\frac{M_0^2}{\sqrt{M_0^2+p^2}}\int \frac{d^3l}{(2\pi)^3}\frac{\vec{\sigma}_1\cdot(\vec{l}+\vec{p}^{\,'})\vec{\sigma}_2\cdot(\vec{l}+\vec{p}^{\,'})\vec{\sigma}_1\cdot(\vec{l}+\vec{p}\,)\vec{\sigma}_2\cdot(\vec{l}+\vec{p}\,)}{(p^2-l^2+i\epsilon)[(\vec{l}+\vec{p}^{\,'})^2+m_2^2][(\vec{l}+\vec{p}\,)^2+m_1^2]}\nonumber\\
=&\frac{M_0^2}{\sqrt{M_0^2+p^2}}\Big\{J_0-\frac{m_1^2+3m_2^2+2q^2}{4}\Gamma_0(m_2)-\frac{m_2^2+3m_1^2+2q^2}{4}\Gamma_0(m_1)\nonumber\\
&-\frac{q^2}{4}[\Gamma_1(m_1)+\Gamma_1(m_2)]+\frac{(m_1^2+m_2^2+q^2)^2}{4}G_0(m_1,m_2)-q^2G_2(m_1,m_2)\vec{\sigma}_1\cdot\vec{\sigma}_2\nonumber\\
&+G_2(m_1,m_2)\vec{\sigma}_1\cdot \vec{q}\,\vec{\sigma}_2\cdot \vec{q}-[\Gamma_0(m_1)+\Gamma_0(m_2)+\Gamma_1(m_1)+\Gamma_2(m_2)\nonumber\\
&-(m_1^2+m_2^2+q^2)G_0(m_1,m_2)-2(m_1^2+m_2^2+q^2)G_1(m_1,m_2)][-i\vec{S}\cdot(\vec{q}\times \vec{k})]\nonumber\\
&-[G_0(m_1,m_2)+4G_1(m_1,m_2)+4G_3(m_1,m_2)]\vec{\sigma}_1\cdot(\vec{q}\times\vec{k})\,\vec{\sigma}_2\cdot(\vec{q}\times\vec{k})\Big\},
\end{align}

where
\begin{align}
\label{j0}
J_0=-\frac{1}{4\pi}ip,
\end{align}
\begin{align}
\label{gamma0}
\Gamma_0(m)=-\frac{1}{8\pi p}(\text{arctan}\frac{2p}{m}+i\text{ln}\frac{\sqrt{m^2+4p^2}}{m}),
\end{align}
\begin{align}
\label{gamma1}
\Gamma_1(m)=-\frac{1}{8\pi p^2}(m+i p)-\frac{1}{2p^2}(m^2+2p^2)\Gamma_0(m),
\end{align}
\begin{align}
\label{g0}
G_0(m_1,m_2)=&-\frac{1}{16\pi}\int^{1}_{0}d x\Big\{\frac{1}{(\delta _{12}^2-4p^2q^2)x^2+2[\delta_{12}(m_2^2+2p^2)+2p^2q^2]x+(m_2^4+4m_2^2p^2)}\nonumber\\
&\times[\frac{2(m_2^2+\delta_{12}x)}{\sqrt{m_2^2+(\delta_{12}+q^2-q^2x)x}}+4ip]\Big\},\quad \delta_{12}=m_1^2-m_2^2,
\end{align}
\begin{align}
\label{g1}
G_1(m_1,m_2)=\frac{1}{4p^2-q^2}\Big\{\frac{1}{2}[\Gamma_0(m_1)+\Gamma_0(m_2)]-(\frac{m_1^2+m_2^2}{2}+2p^2)G_0(m_1,m_2)+\frac{A(m_1,m_2)}{2\pi}\Big\},
\end{align}
\begin{align}
\label{g2}
G_2(m_1,m_2)=p^2G_0(m_1,m_2)+(\frac{m_1^2+m_2^2}{2}+2p^2)G_1(m_1,m_2)-\frac{A(m_1,m_2)}{4\pi},
\end{align}
\begin{align}
\label{g3}
G_3(m_1,m_2)=\frac{1}{4p^2-q^2}\Big\{\frac{1}{4}[\Gamma_1(m_1)+\Gamma_2(m_2)]-p^2G_0(m_1,m_2)-(m_1^2+m_2^2+4p^2)G_1(m_1,m_2)\Big\},
\end{align}
with $A(m_1,m_2)$ given in Eq.~(\ref{Afunction}). In terms of the integral function $V_{\text{it}}(m_1,m_2)$, the iterated one-meson-exchange contributions to the nucleon-nucleon amplitudes read
\begin{align}
\label{Vit}
V^{\text{it}}_{(C,S,T,LS,\sigma L)}=\frac{3g_A^4}{16f_\pi^4}V_{\text{it}}(m_\pi,m_\pi)+\frac{g_\eta^4}{144 f_\eta^4}V_{\text{it}}(m_\eta,m_\eta),
\end{align}
\begin{align}
\label{Wit}
W^{\text{it}}_{(C,S,T,LS,\sigma L)}=-\frac{g_A^4}{8f_\pi^4}V_{\text{it}}(m_\pi,m_\pi)+\frac{g_A^2g_\eta^2}{48f_\eta^2f_\pi^2}[V_{\text{it}}(m_\eta,m_\pi)+V_{\text{it}}(m_\pi,m_\eta)].
\end{align}
It is straightforward to obtain the central, spin-spin, tensor, spin-orbit and quadratic spin-orbit components of the $T$ matrix. Note that, the occurrence of the factor $M_0$ in iterated diagrams changes the correspondence between the loop expansion and the small momentum expansion for processes involving two nucleons. For the above iterated one-meson exchange, the power counting is two because the number of the loop is one. The actual contribution from the iterated one-meson exchange to the nucleon-nucleon $T$ matrix is of first order in small momenta, even though we count the iterated one-meson-exchange diagram as next-to-leading order in order to obtain a consistent result with the SU(2) case and make a comparison. Besides, the further iterations correspond to the diagrams with many more loops which are extremely difficult to evaluate. Numerically, we find that in most cases the iterated one-meson-exchange contribution is small compared with the irreducible meson-exchange contribution. Therefore, we assume that the higher-order iterations can be neglected.

\section{Phase shifts and mixing angles}
\label{phaseshifts}
In order to calculate phase shifts and mixing angles, the matrix elements of $T(\vec{p}\,',\vec{p})$ in the $LSJ$ basis are needed, where $S$ denotes the total spin, $L$ the total orbital angular momentum, and $J$ the total angular momentum. The $T$ matrix is decomposed into partial waves following Ref.~\cite{erke1971}, and the following projection formulas are obtained:\\
(a) Spin singlet with $S=0$ and $L=J$:
\begin{align}
\label{S0LJ}
\langle J0J|T|J0J \rangle=\frac{1}{2}\int^{1}_{-1}dz[U_C-3U_S-q^2U_T+p^4(z^2-1)U_{\sigma L}]P_J(z).
\end{align}
(b) Uncoupled spin triplet with $S=1$ and $L=J$:
\begin{align}
\label{S1LJ}
\langle J1J|T|J1J \rangle=&\frac{1}{2}\int^{1}_{-1}dz\{2p^2(-U_{LS}/2-U_T+p^2zU_{\sigma L})[P_{J+1}(z)+P_{J-1}(z)]+[U_C+U_S\nonumber\\
&+2p^2(1+z)U_T+2p^2zU_{LS}-p^4(3z^2+1)U_{\sigma L}]P_J(z)\}.
\end{align}
(c)Coupled triplet states with $S=1$ and $L=J\pm 1$:
\begin{align}
\label{S1LJpm1}
\langle J\pm 1,1J|T|J\pm 1, 1J \rangle=&\frac{1}{2}\int^{1}_{-1}dz\{2p^2[-U_{LS}/2\pm \frac{1}{2J+1}(U_T-p^2zU_{\sigma L})]P_J(z)+[U_C+U_S\nonumber\\
&+p^2(p^2(1-z^2)U_{\sigma L}+z U_{LS}\pm \frac{2}{2J+1}(p^2U_{\sigma
L}-U_T))]P_{J\pm 1}(z)\}.
\end{align}
(d)Coupled triplet states with $S=1$, $L^{'}=J-1$, and $L=J+1$:
\begin{align}
\label{S1LJadd1}
\langle J-1,1J|T|J+1, 1J \rangle=&\frac{\sqrt{J+1}p^2}{\sqrt{J}(2J+1)}\int^{1}_{-1}dz\{(U_T-p^2U_{\sigma L})P_{J+1}(z)\nonumber\\
&+[(2J-z(2J+1))U_T+p^2zU_{\sigma L}]P_J(z)\}.
\end{align}
Here, $P_J(z)$ are ordinary Legendre polynomials of degree $J$. The $U_{C,...,\sigma L}$ are given by
\begin{align}
\label{UK}
U_K=V_K+(4I-3)W_K,\quad\quad (K=C,S,T,LS,\sigma L),
\end{align}
with total isospin $I=0,1$. The $L+S+I$ must be odd because of the Pauli exclusion principle.

The phase shifts and mixing angles are calculated via
\begin{align}
\label{delta}
\delta_{LSJ}=-\frac{M_N^2p}{4\pi E_N}\text{Re}\langle LSJ|T|LSJ\rangle,
\end{align}
\begin{align}
\label{epsilon}
\epsilon_{J}=-\frac{M_N^2p}{4\pi E_N}\text{Re}\langle J-1,1J|T|J+1,1J\rangle,
\end{align}
with the c.m.s. nucleon energy $E_N=\sqrt{M_N^2+p^2}$. Note that, the calculation of the phase shift and the mixing angle based on Eqs.~(\ref{delta}) and (\ref{epsilon}) is valid only as long as the difference between $\delta$ and $\sin\delta\cos\delta$ is small. In addition, the c.m.s. momentum $p$ is related to the kinetic energy of the incident neutron in the laboratory system $T_{\text{lab}}$ by
\begin{align}
\label{psqur}
p^2=\frac{M_p^2T_{\text{lab}}(T_{\text{lab}}+2M_n)}{(M_p+M_n)^2+2T_{\text{lab}}M_p}.
\end{align}

\section{Results and discussion}
\label{results}
In this section, for the $T$ matrix given in Eq.~(\ref{tmatrice}), we present and discuss our results for the phase shifts with $2 \leq L \leq 6$ and mixing angles with $2 \leq J \leq 6$ up to $T_{\text{lab}}=250 \,\text{MeV}$. At such an energy, momentum transfers up to $q=4.9\,m_\pi=685\,\text{MeV}$ are involved, which are still quite large for the application of chiral perturbation theory. Nonetheless, we give results up to 250 MeV in order to show where our predictions are consistent with the existing data and where deviations appear. Before presenting the results, we summarize the values of the physical parameters: $m_\pi=139.57\,\text{MeV},\,m_K=493.68\,\text{MeV},\,m_\eta=547.86\,\text{MeV},\,f_{\pi}=92.07\,\text{MeV},\,f_{K}=110.03\,\text{MeV},\,f_\eta=1.2f_\pi,\,M_p=938.27\,\text{MeV},\,M_n=939.57\,\text{MeV},\,M_N=938.92\,\text{MeV},\,\lambda=4\pi f_\pi=1.16\text{GeV},\,D=0.80,$ and $F=0.47\,$\cite{pdg2018,chan2018,mark2019}. We use $M_0=963.58\pm 153.97\,\text{MeV},\,b_D=0.06\,\text{GeV}^{-1},\,b_F=-0.48\,\text{GeV}^{-1},\,b_0=-0.69\,\text{GeV}^{-1}$ from the results of the pion-nucleon scattering at chiral order $\mathcal{O}(p^3)$ in our previous paper \cite{huan20201}. Unfortunately, the remaining low-energy constants $b_{i=1,...,11}$ were combined into three linear combinations $C_{1,2,3}$. The separated $b_{i}$ were not obtained at this order. However, the respective values of the $b_{i=1,...,11}$ have been obtained in pion-nucleon scattering at order $p^4$ in our other paper \cite{huan20202}. We take the low-energy constants $b_{i=1,...,11}$ used in the calculation: $b_1=1.61\,\text{GeV}^{-1},\,b_2=0.10\,\text{GeV}^{-1},\,b_3=-4.50\,\text{GeV}^{-1},\,b_4=-1.34\,\text{GeV}^{-1},\,b_9=1.86\,\text{GeV}^{-1},\,b_{10}=-0.38\,\text{GeV}^{-1},\,b_{11}=17.66\,\text{GeV}^{-1}$. This may cause some errors. Thus, we give a common uncertainty of $\pm 20 \%$ to the low-energy constants $b_{D,F,0,...,4,9,10,11}$. The error bands will be generated by the standard error propagation formula, $\delta^2 \mathcal{O}=(\partial \mathcal{O}/\partial x_1)^2(\delta x_1)^2+(\partial \mathcal{O}/\partial x_2)^2(\delta x_2)^2+...$, using the uncertainties of the low-energy constants $b_i$ and the chiral limit baryon mass $M_0$ without correlations. Although the errors are crude, they can measure how dependent the phase shifts and the mixing angles are on the $b_i$ and $M_0$.
Furthermore, the orders displayed will be defined as LO [one-meson exchange, Eq.~(\ref{VWLO})], NLO [LO plus iterated one-meson exchange, Eqs.~(\ref{Vit}) and (\ref{Wit}), plus the contributions of second order, Eqs.~(\ref{VCNLO})~(\ref{WTWSNLO})] and NNLO [NLO plus the contributions of third order, Eqs.~(\ref{VCNNLO})~(\ref{WLSNNLO})] in following figures.

\begin{figure}[!t]
\centering
\includegraphics[height=16cm,width=12cm]{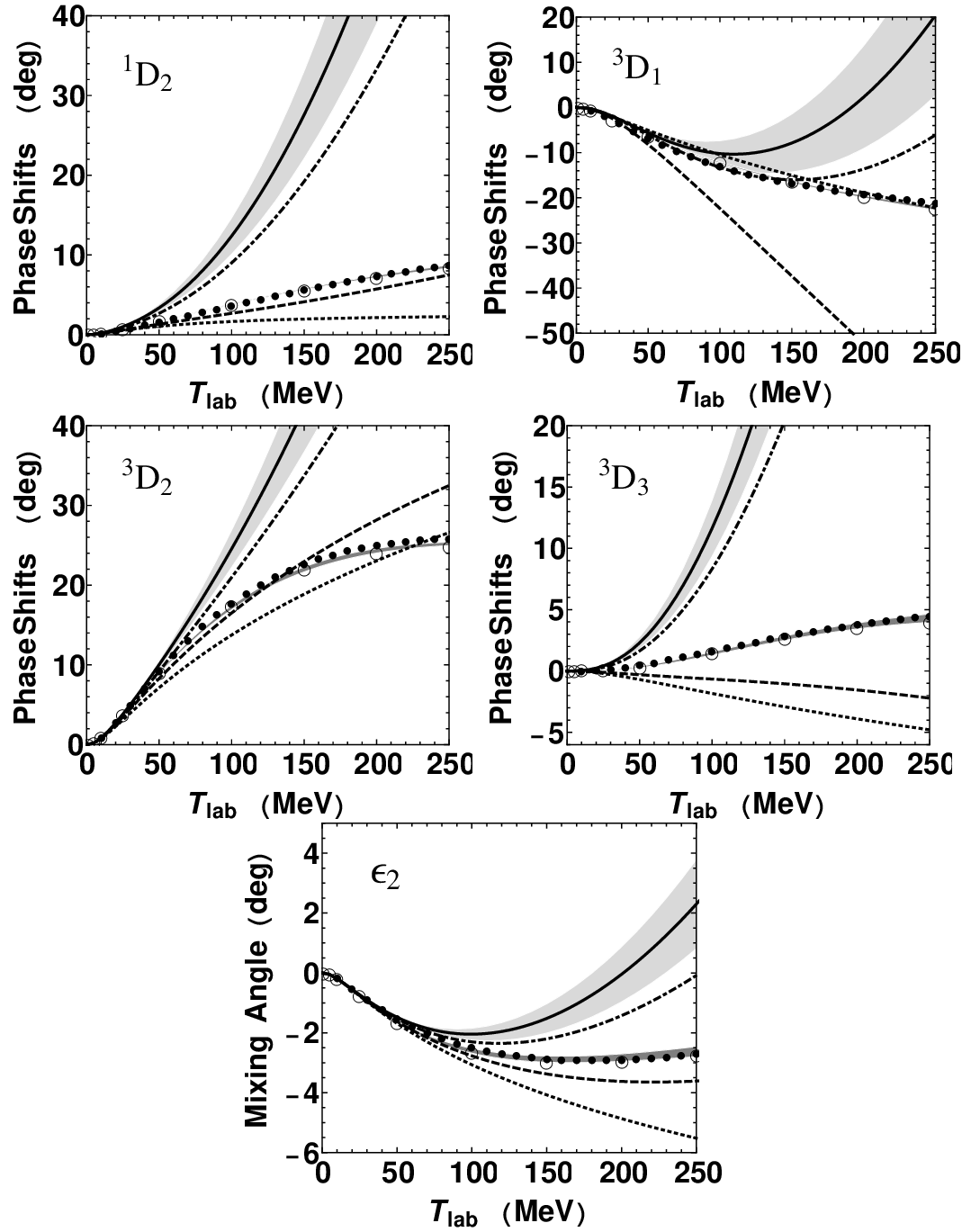}
\caption{\label{fig:dwaves}D-wave $NN$ phase shifts and mixing angle $\epsilon_2$ versus the nucleon laboratory kinetic energy $T_{\text{lab}}$. The dotted, dashed and solid lines present the predictions to LO, NLO, NNLO, respectively. The dot-dashed lines refer to the results from the SU(2) H$\chi$PT at order NNLO, the black dots denote the SM16 solutions, and the open circles present the Granada 2013 results. The bands (light gray) are generated by the standard error propagation formula with the errors of the low-energy constants $b_i$ and the chiral limit baryon mass $M_0$, and the bands (gray) including 13 high statistical quality potentials are given at $1\sigma$ confidence level. See main text. }
\end{figure}

\begin{figure}[!t]
\centering
\includegraphics[height=16cm,width=12cm]{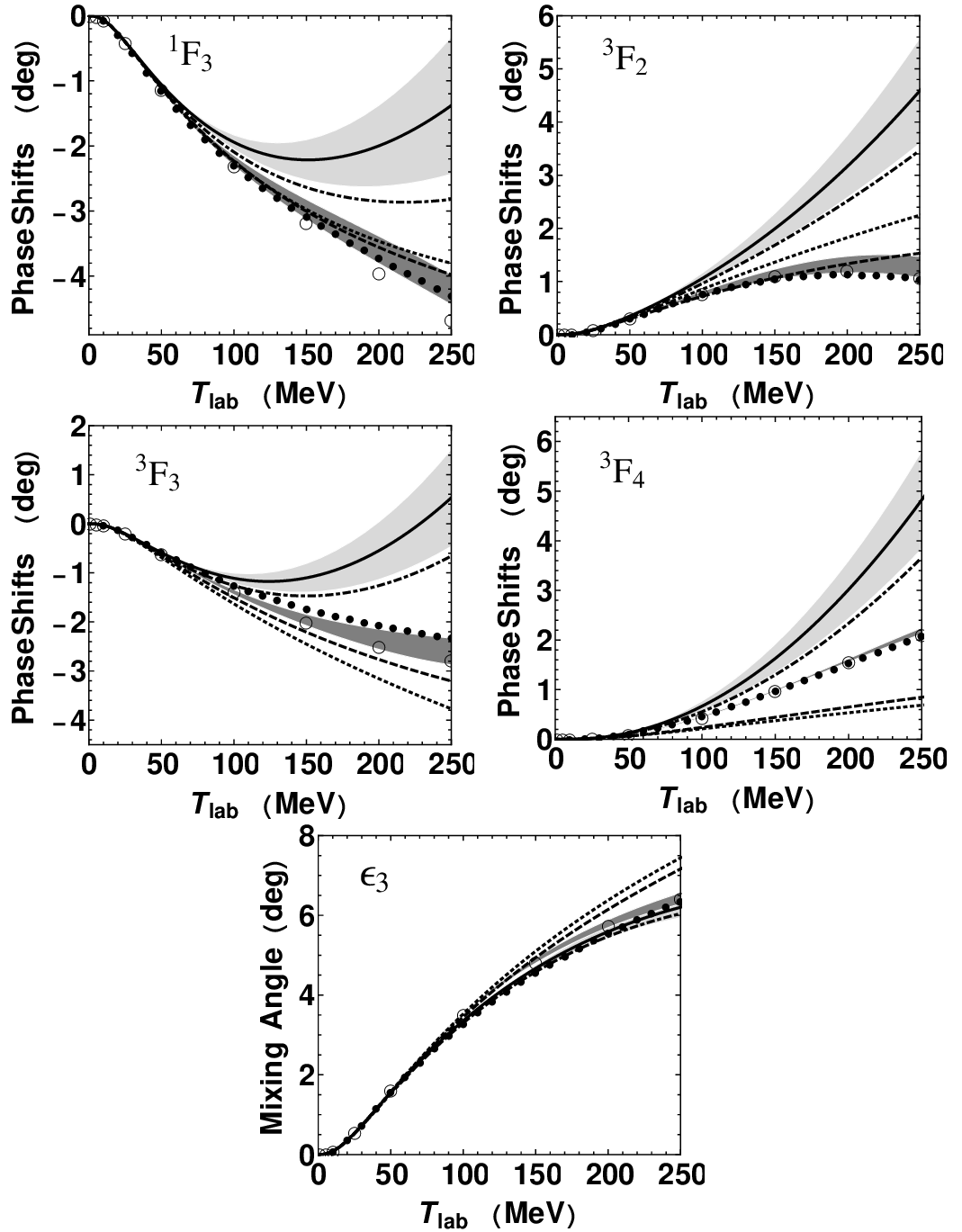}
\caption{\label{fig:fwaves}F-wave $NN$ phase shifts and mixing angle $\epsilon_3$ versus the nucleon laboratory kinetic energy $T_{\text{lab}}$. For notation, see Fig.~\ref{fig:dwaves}.}
\end{figure}

\begin{figure}[!t]
\centering
\includegraphics[height=16cm,width=12cm]{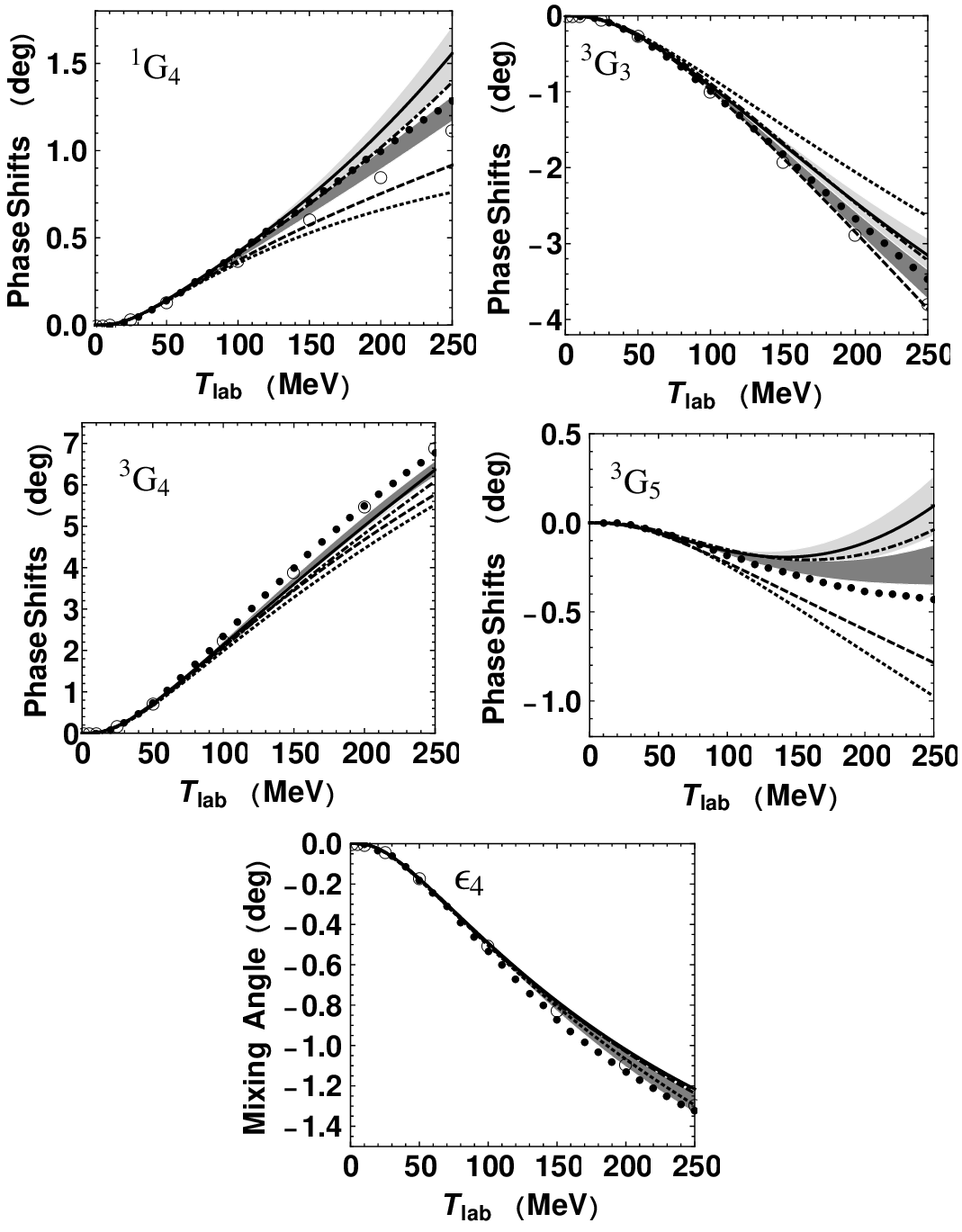}
\caption{\label{fig:gwaves}G-wave $NN$ phase shifts and mixing angle $\epsilon_4$ versus the nucleon laboratory kinetic energy $T_{\text{lab}}$. For notation, see Fig.~\ref{fig:dwaves}.}
\end{figure}

\begin{figure}[!t]
\centering
\includegraphics[height=16cm,width=12cm]{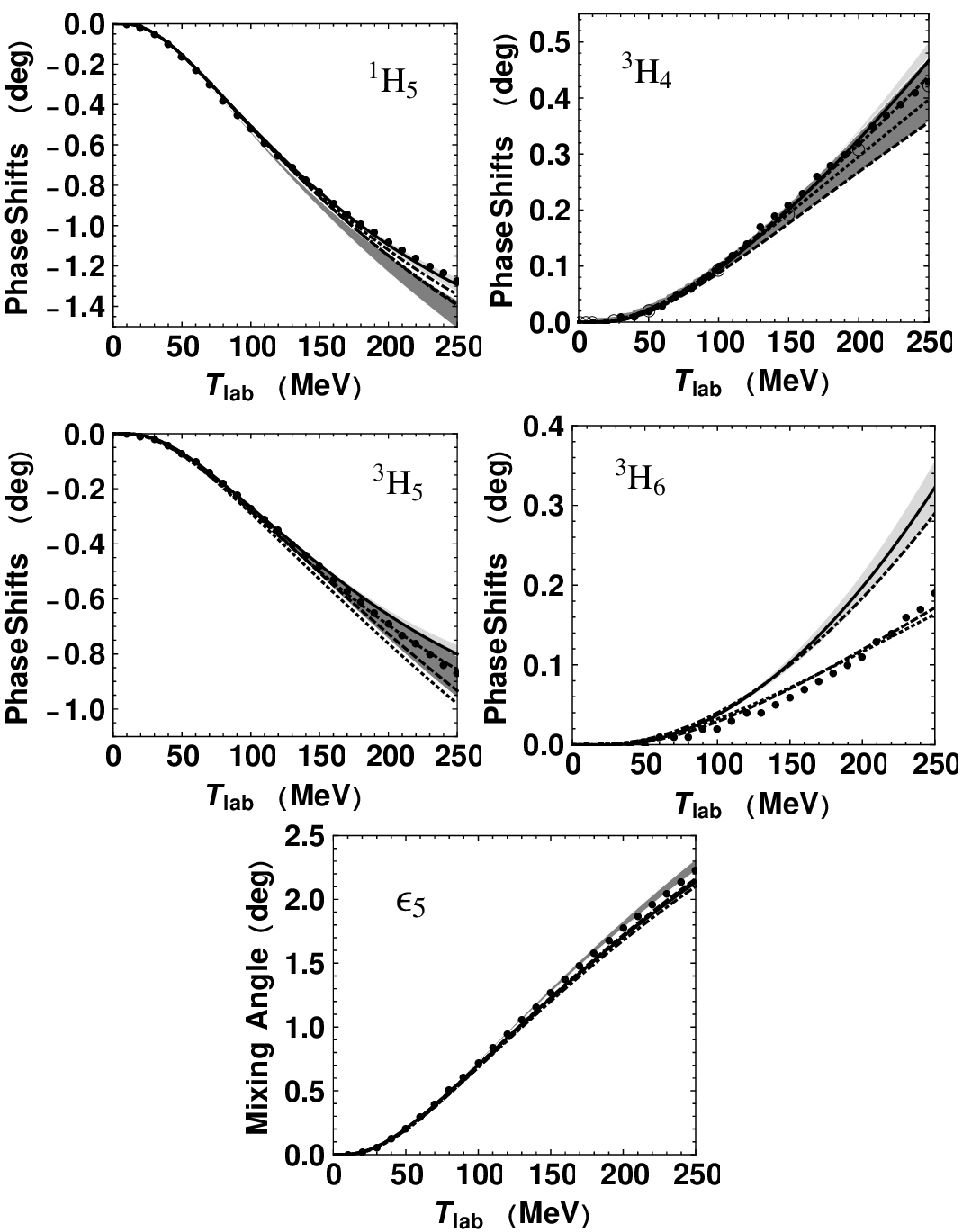}
\caption{\label{fig:hwaves}H-wave $NN$ phase shifts and mixing angle $\epsilon_5$ versus the nucleon laboratory kinetic energy $T_{\text{lab}}$. For notation, see Fig.~\ref{fig:dwaves}. }
\end{figure}

\begin{figure}[!t]
\centering
\includegraphics[height=16cm,width=12cm]{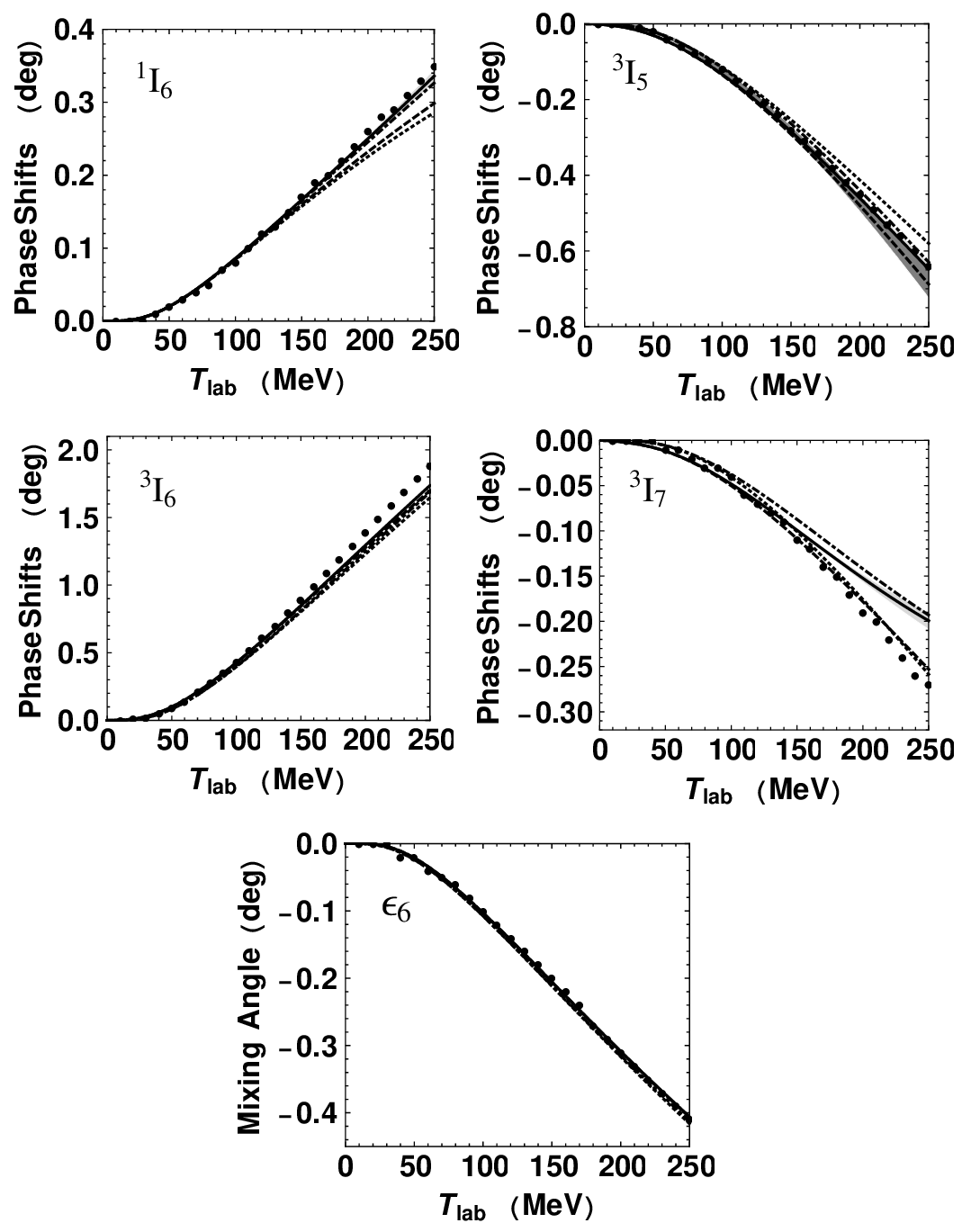}
\caption{\label{fig:iwaves}I-wave $NN$ phase shifts and mixing angle $\epsilon_6$ versus the nucleon laboratory kinetic energy $T_{\text{lab}}$. For notation, see Fig.~\ref{fig:dwaves}.}
\end{figure}

\subsection{D wave}
The D wave phase shifts and the mixing angle $\epsilon_2$ are shown in Fig.~\ref{fig:dwaves}. The dotted lines correspond to the one-meson-exchange approximation, i. e., LO. The dashed and solid lines present the NLO and NNLO, respectively. The black dots denote the SM16 solutions from SAID online \citep{SAID}. The open circles present the Granada 2013 results from the upgrade of the nucleon-nucleon database by the Granada group\cite{nava20131,nava20132}. The results from Granada 2013 have uncertainties, but we do not show them because the uncertainties are small, below 250 MeV laboratory kinetic energy; for more details about the analysis, see Refs.~\cite{nava20201,nava20202}. The dot-dashed lines refer to the results from the SU(2) HB$\chi$PT at order NNLO \cite{kais19971}. The light gray bands are generated by the standard error propagation formula with the errors of the low-energy constants $b_i$ and the chiral limit baryon mass $M_0$. The gray bands including 13 high statistical quality potentials are given at $1\sigma$ confidence level from Fig. 10 of Ref.~\cite{simo2018}. These potentials include the Nijmegen PWA \cite{stok1993}, NijmI, NijmII, Reid93 \cite{stok1994}, AV18 \cite{wiri1995}, CD Bonn \cite{mach2001}, Spectator \cite{gros2008}, and the six Granada potentials denoted as DS-OPE \cite{nava20131,nava20132}, DS-$\chi$TPE \cite{nava20141,nava20142}, SOG-OPE \cite{nava20143}, SOG-$\chi$TPE, DS-$\Delta$BO, and SOG-$\Delta$BO \cite{nava2016}. In all cases, the chiral nucleon-nucleon phase shifts at third order go into the proper direction, that means the two-meson-exchange corrections give consistent signs of the phase shifts with empirical values below 100 MeV laboratory kinetic energy. But they are too large to keep the signs of the phase shifts at high energies in some cases and even larger than the SU(2) case. Not surprisingly, the coupling constants $b_i$ in our calculation have large errors. When considering the bands from the errors of the constants, our results are almost consistent with the SU(2) case. However, our predictions for the D-wave phase shifts obviously deviate from the data. One exception is the mixing angle $\epsilon_2$ which is consistent with the data below 100 MeV laboratory kinetic energy. The leading-order two-meson-exchange contributions are not small for the D-wave phase shifts, especially for $^{3}D_{1}$ phase shifts. This is different from the SU(2) case. This also results in the $^{3}D_{1}$ phase shift being only in fair agreement with the data up to 100 MeV laboratory kinetic energy. In the SU(2) case, there exists an almost complete cancellation of irreducible two-pion exchange and iterated one-pion exchange contributions for the $^{3}D_{1}$ wave. Nevertheless, for the SU(3) case, the iterated one-meson-exchange contributions only include the $\pi$ and $\eta$-meson exchange, while the irreducible two-meson-exchange contributions involve $\pi$, $K$, and $\eta$-meson exchange. Thus, this cancellation mechanism does not work anymore in the SU(3) case. The one-meson-exchange contribution in the $^{3}D_{1}$ wave is roughly consistent with the data. The situation is same as the SU(2) case. This shows that the $\eta$-meson-exchange contribution is small. At last, we also do not obtain a proper convergence. To sum up, the one-meson and two-meson exchange alone is also not sufficient to describe the dynamics in the nucleon-nucleon D waves, as concluded in the SU(2) case \cite{kais19971}.

\subsection{F wave}
The F-wave phase shifts and the mixing angle $\epsilon_3$ are shown in Fig.~\ref{fig:fwaves}. Again, our predictions go into the right direction and are roughly consistent with the SU(2) case when the error bands are considered. The predictions for the phase shifts in the $^1F_3$, $^3F_2$, $^3F_3$, and $^3F_4$ partial waves are in good agreement with the data up to 100 MeV laboratory kinetic energy. There are also large error bands in the four waves, especially for $^1F_3$. This means that we can improve the description of the four F waves by refitting the data. The mixing angle $\epsilon_3$ with a small error band, however, is in perfect agreement with the data for all energies up to 250 MeV laboratory kinetic energy, the same as in the SU(2) case. Our prediction at third order is almost consistent with the result from the one-meson-exchange contribution below 150 MeV for the mixing angle $\epsilon_3$. The leading-order two-meson-exchange contributions are small for the F-wave phase shifts and the mixing angle $\epsilon_3$. The situation is the same as in the SU(2) case. The convergence is still not appropriate for all four partial-wave phase shifts and the mixing angle at high energies. All in all, the dynamics in the SU(3) case for the nucleon-nucleon F waves and the mixing angle $\epsilon_3$ is almost the same as in the SU(2) case. The predictions for F-wave phase shifts are also clearly dependent on the low-energy constants. This can be used to refit the data to improve the description of the nucleon-nucleon phase shifts and obtain more accurate low-energy constants in the future.

\subsection{G wave}
The G-wave phase shifts and the mixing angle $\epsilon_4$ are shown in Fig.~\ref{fig:gwaves}. Our predictions are in good agreement with data for all partial-wave phase shifts (except for $^{3}G_5$) and the mixing angle up to 250 MeV laboratory kinetic energy within error. The phase shifts for $^{3}G_5$ are only consistent with the data below 150 MeV laboratory kinetic energy. When considering the error band, the description of the $^{3}G_5$ wave cannot be improved. The problem can be solved when the higher-order contributions are considered. The differences between one-meson exchange and the data are still sizeable in the $^{1}G_4$ and $^{3}G_5$ partial waves. For the $^{3}G_3$ and $^{3}G_4$ phase shifts and $\epsilon_4$, the two-meson-exchange corrections are small. In short, our predictions for the four partial-wave phase shifts and the mixing angle are almost the same as in the SU(2) case.

\subsection{H wave}
The H-wave phase shifts and the mixing angle $\epsilon_5$ are shown in Fig.~\ref{fig:hwaves}. The contributions from two-meson exchange are quite small. Our predictions are also in good agreement with data for all partial-wave phase shifts (except for $^{3}H_6$) and the mixing angle up to 250 MeV laboratory kinetic energy. The $^{3}H_6$ phase shifts come out too large. Nevertheless, the $^{3}H_6$ phase shifts are very small, less than $0.5^{\circ}$, and the gap between the data and our prediction may be insignificant if one considers the nucleon-nucleon scattering data. We also find that the contributions from one $\eta$-meson exchange are also very small in the four  phase shifts and the mixing angle, but still obviously make some difference. Once again, our results are the same as in the SU(2) case in the phase shifts and the mixing angle.

\subsection{I wave}
The I-wave phase shifts and the mixing angle $\epsilon_6$ are shown in Fig.~\ref{fig:iwaves}. The two-meson-exchange contributions are still quite small, and our predictions are in very good agreement with the data for all partial-wave phase shifts and the mixing angle up to 250 MeV laboratory kinetic energy. There is only a little gap in $^{3}I_7$ at high energy, and it also is insignificant because the phase shifts for this wave are very small. The contributions from $\eta$-meson exchange are quite small and can even be completely ignored. Furthermore, we find fast convergence in the high angular momentum partial waves (except for $^{3}I_7$) as the existing nucleon-nucleon phase shifts analyses.

\section{Summary}
\label{summary}
We have calculated the complete $T$ matrices of elastic nucleon-nucleon scattering up to third order in SU(3) HB$\chi$PT. With the $T$ matrices, the phase shifts with orbital angular momentum $L\geq2$ and the mixing angles with $J\geq 2$ have been evaluated by using low-energy constants that were extracted from meson-baryon analysis. Then, we have compared our predictions with the empirical phase shifts and mixing angles data, and as well as the results from SU(2) HB$\chi$PT. The description of the data within the SU(3) case is slightly worse than the SU(2) case because a large number of low-energy constants cannot be determined accurately at this order. In addition, we have found that one-meson and two-meson exchange alone is not sufficient to describe the dynamics in the nucleon-nucleon D waves, the same as in the SU(2) case. For higher partial waves, the model-independent amplitudes up to third order in the SU(3) case bring the chiral prediction close to empirical nucleon-nucleon phase shifts, as in the SU(2) case. Meanwhile, in the high angular momentum partial waves($L\geq5$), we find fast convergence to the result of one-meson exchange as obtained in the empirical nucleon-nucleon phase shifts analyses. The errors of phase shifts and mixing angles have been obtained by the standard error propagation formula with the errors of the low-energy constants $b_i$ and the chiral limit baryon mass $M_0$. Besides D waves, the predictions for F-wave phase shifts were also clearly dependent on the low-energy constants. This can be used to refit the data to improve the description of the nucleon-nucleon phase shifts and obtain the more accurate low-energy constants in the future. In a word, our predictions for the peripheral nucleon-nucleon phase scattering in SU(3) HB$\chi$PT are quite reasonable. The higher-order amplitudes for nucleon-nucleon scattering and much more baryon-baryon scattering channels can be investigated further in the SU(3) case.
\section*{Acknowledgements}
This work is supported by the National Natural Science Foundation of China under Grants No.11975033, No.12070131001, and No.12147127, and the China Postdoctoral Science Foundation (Grant No.2021M700251). We thank Norbert Kaiser (Technische Universit\"{a}t M\"{u}nchen), Yan-Rui Liu (Shandong University), and Zi-Yang Lin (Peking University) for very helpful discussions.

\bibliographystyle{unsrt}
\bibliography{latextemplate}

\providecommand{\noopsort}[1]{}\providecommand{\singleletter}[1]{#1}%
\begin{thebibliography}{10}

\bibitem{wein1979}
S.~Weinberg.
\newblock {Phenomenological Lagrangians}.
\newblock {\em Physica A}, 96:327--340, 1979.

\bibitem{sche2012}
S.~Scherer and M.~R. Schindler.
\newblock {A primer for chiral perturbation theory}.
\newblock {\em Lect. Notes Phys.}, 830:pp.1--338, 2012.

\bibitem{jenk1991}
E.~E. Jenkins and A.~V. Manohar.
\newblock {Baryon chiral perturbation theory using a heavy fermion Lagrangian}.
\newblock {\em Phys. Lett. B}, 255:558--562, 1991.

\bibitem{bern1992}
V.~Bernard, N.~Kaiser, J.~Kambor, and U.-G. Mei{\ss}ner.
\newblock {Chiral structure of the nucleon}.
\newblock {\em Nucl. Phys. B}, 388:315--345, 1992.

\bibitem{bech1999}
T.~Becher and H.~Leutwyler.
\newblock {Baryon chiral perturbation theory in manifestly Lorentz invariant
  form}.
\newblock {\em Eur. Phys. J. C}, 9:643--671, 1999.

\bibitem{gege1999}
J.~Gegelia and G.~Japaridze.
\newblock {Matching heavy particle approach to relativistic theory}.
\newblock {\em Phys. Rev. D}, 60:114038, 1999.

\bibitem{fuch2003}
T.~Fuchs, J.~Gegelia, G.~Japaridze, and S.~Scherer.
\newblock {Renormalization of relativistic baryon chiral perturbation theory
  and power counting}.
\newblock {\em Phys. Rev. D}, 68:056005, 2003.

\bibitem{schi2007}
M.~R. Schindler, T.~Fuchs, J.~Gegelia, and S.~Scherer.
\newblock {Axial, induced pseudoscalar, and pion-nucleon form-factors in
  manifestly Lorentz-invariant chiral perturbation theory}.
\newblock {\em Phys. Rev. C}, 75:025202, 2007.

\bibitem{geng2008}
L.~S. Geng, J.~Martin~Camalich, L.~Alvarez-Ruso, and M.~J. Vicente~Vacas.
\newblock {Leading SU(3)-breaking corrections to the baryon magnetic moments in
  chiral perturbation theory}.
\newblock {\em Phys. Rev. Lett.}, 101:222002, 2008.

\bibitem{mart2010}
J.~Martin~Camalich, L.~S. Geng, and M.~J. Vicente~Vacas.
\newblock {The lowest-lying baryon masses in covariant SU(3)-flavor chiral
  perturbation theory}.
\newblock {\em Phys. Rev. D}, 82:074504, 2010.

\bibitem{ren2012}
X.~L. Ren, L.~S. Geng, J.~Martin~Camalich, J.~Meng, and H.~Toki.
\newblock {Octet baryon masses in next-to-next-to-next-to-leading order
  covariant baryon chiral perturbation theory}.
\newblock {\em JHEP}, 12:073, 2012.

\bibitem{alar2012}
J.~M. Alarc\'{o}n, J.~Martin~Camalich, and J.~A. Oller.
\newblock {The chiral representation of the $\pi N$ scattering amplitude and
  the pion-nucleon sigma term}.
\newblock {\em Phys. Rev. D}, 85:051503, 2012.

\bibitem{alar2013}
J.~M. Alarc\'{o}n, J.~Martin~Camalich, and J.~A. Oller.
\newblock {Improved description of the $\pi N$ scattering phenomenology in
  covariant baryon chiral perturbation theory}.
\newblock {\em Annals Phys.}, 336:413--461, 2013.
\newblock [Private communication with J.M. Alarc\'{o}n].

\bibitem{chen2013}
Y.-H. Chen, D.-L. Yao, and H.~Q. Zheng.
\newblock {Analyses of pion-nucleon elastic scattering amplitudes up to
  $\mathcal{O}(p^4)$ in extended-on-mass-shell subtraction scheme}.
\newblock {\em Phys. Rev. D}, 87:054019, 2013.

\bibitem{yao2016}
D.-L. Yao, D.~Siemens, V.~Bernard, E.~Epelbaum, A.~M. Gasparyan, J.~Gegelia,
  H.~Krebs, and U.-G. Mei{\ss}ner.
\newblock {Pion-nucleon scattering in covariant baryon chiral perturbation
  theory with explicit Delta resonances}.
\newblock {\em JHEP}, 05:038, 2016.

\bibitem{lu2019}
J.-X. Lu, L.-S. Geng, X.-L. Ren, and M.-L. Du.
\newblock {Meson-baryon scattering up to the next-to-next-to-leading order in
  covariant baryon chiral perturbation theory}.
\newblock {\em Phys. Rev. D}, 99(5):054024, 2019.

\bibitem{wein1990}
S.~Weinberg.
\newblock {Nuclear forces from chiral lagrangians}.
\newblock {\em Phys. Lett. B}, 251:288--292, 1990.

\bibitem{wein1991}
S.~Weinberg.
\newblock {Effective chiral lagrangians for nucleon-pion interations and
  nuclear forces}.
\newblock {\em Nucl. Phys. B}, 363:3--18, 1991.

\bibitem{ordo1992}
C.~Ordóñez and U.~{van Kolck}.
\newblock Chiral lagrangians and nuclear forces.
\newblock {\em Phys. Lett. B}, 291(4):459--464, 1992.

\bibitem{cele1992}
L.~S. Celenza, A.~Pantziris, and C.~M. Shakin.
\newblock Chiral symmetry and the nucleon-nucleon interaction: Tensor
  decomposition of feynman diagrams.
\newblock {\em Phys. Rev. C}, 46:2213--2223, Dec 1992.

\bibitem{ordo1994}
C.~Ord\'o\~nez, L.~Ray, and U.~van Kolck.
\newblock Nucleon-nucleon potential from an effective chiral lagrangian.
\newblock {\em Phys. Rev. Lett.}, 72:1982--1985, Mar 1994.

\bibitem{ball1998}
J-L. Ballot, M.~R. Robilotta, and C.~A. da~Rocha.
\newblock $\mathrm{NN}$ scattering: Chiral predictions for asymptotic
  observables.
\newblock {\em Phys. Rev. C}, 57:1574--1579, Apr 1998.

\bibitem{kais1998}
N.~Kaiser, S.~Gerstendörfer, and W.~Weise.
\newblock Peripheral {NN}-scattering: role of delta-excitation, correlated
  two-pion and vector meson exchange.
\newblock {\em Nucl. Phys. A}, 637(3):395--420, 1998.

\bibitem{epel1998}
E.~Epelbaoum, W.~Glöckle, and Ulf-G. Meißner.
\newblock Nuclear forces from chiral lagrangians using the method of unitary
  transformation (i): Formalism.
\newblock {\em Nucl. Phys. A}, 637(1):107--134, 1998.

\bibitem{kais2001}
N.~Kaiser.
\newblock {Chiral corrections to kaon nucleon scattering lengths}.
\newblock {\em Phys. Rev. C}, 64:045204, 2001.
\newblock [Erratum: Phys. Rev.C73,069902(2006)].

\bibitem{ente2002}
D.R. Entem and R.~Machleidt.
\newblock Accurate nucleon–nucleon potential based upon chiral perturbation
  theory.
\newblock {\em Phys. Lett. B}, 524(1):93--98, 2002.

\bibitem{ente2015}
D.~R. Entem, N.~Kaiser, R.~Machleidt, and Y.~Nosyk.
\newblock {Peripheral nucleon-nucleon scattering at fifth order of chiral
  perturbation theory}.
\newblock {\em Phys. Rev. C}, 91(1):014002, 2015.

\bibitem{wu2018}
Shaowei Wu and Bingwei Long.
\newblock {Perturbative $NN$ scattering in chiral effective field theory}.
\newblock {\em Phys. Rev. C}, 99(2):024003, 2019.

\bibitem{kais2019}
N.~Kaiser.
\newblock {Density-dependent NN interaction from subsubleading chiral 3N
  forces: Intermediate-range contributions}.
\newblock {\em Phys. Rev. C}, 101(1):014001, 2020.

\bibitem{higa2003}
R.~Higa and M.~R. Robilotta.
\newblock {Two pion exchange nucleon nucleon potential: Relativistic chiral
  expansion}.
\newblock {\em Phys. Rev. C}, 68:024004, 2003.

\bibitem{higa2004}
R.~Higa, M.~R. Robilotta, and C.~A. da~Rocha.
\newblock {Relativistic O($q^4$) two pion exchange nucleon-nucleon potential:
  Configuration space}.
\newblock {\em Phys. Rev. C}, 69:034009, 2004.

\bibitem{higa2008}
R.~Higa, M.~Pavon~Valderrama, and E.~Ruiz~Arriola.
\newblock {Renormalization of NN Interaction with Relativistic Chiral Two Pion
  Exchange}.
\newblock {\em Phys. Rev. C}, 77:034003, 2008.

\bibitem{ren2018}
X.-L. Ren, K.-W. Li, L.-S. Geng, B.~Long, P.~Ring, and J.~Meng.
\newblock Leading order relativistic chiral nucleon-nucleon interaction.
\newblock {\em Chin. Phys. C}, 42(1):014103, 2018.

\bibitem{xiao2020}
Y.~Xiao, C.-X. Wang, J.-X Lu, and L.-S. Geng.
\newblock {Two-pion exchange contributions to the nucleon-nucleon interaction
  in covariant baryon chiral perturbation theory}.
\newblock {\em Phys. Rev. C}, 102:054001, 2020.

\bibitem{ren2021}
X.-L. Ren, C.-X. Wang, K.-W. Li, L.-S. Geng, and J.~Meng.
\newblock Relativistic chiral description of the $^{1}s_{0}$ nucleon–nucleon
  scattering.
\newblock {\em Chin. Phys. Lett.}, 38(6):062101, 2021.

\bibitem{haid2013}
J.~Haidenbauer, S.~Petschauer, N.~Kaiser, U.-G. Meißner, A.~Nogga, and
  W.~Weise.
\newblock Hyperon–nucleon interaction at next-to-leading order in chiral
  effective field theory.
\newblock {\em Nucl. Phys. A}, 915:24--58, 2013.

\bibitem{haid2020}
J.~Haidenbauer, U.-G. Meißner, and A.~Nogga.
\newblock Hyperon–nucleon interaction within chiral effective field theory
  revisited.
\newblock {\em Eur. Phys. J. A}, 56:91, 2020.

\bibitem{huan2015}
B.-L. Huang and Y.-D. Li.
\newblock {Kaon-nucleon scattering to one-loop order in heavy baryon chiral
  perturbation theory}.
\newblock {\em Phys. Rev. D}, 92(11):114033, 2015.
\newblock [Erratum: Phys. Rev.D95,019903(2017)].

\bibitem{huan2017}
B.-L. Huang, J.-S. Zhang, Y.-D. Li, and N.~Kaiser.
\newblock {Meson-baryon scattering to one-loop order in heavy baryon chiral
  perturbation theory}.
\newblock {\em Phys. Rev. D}, 96(11):016021, 2017.

\bibitem{huan20201}
B.-L. Huang and J.~Ou-Yang.
\newblock {Pion-nucleon scattering to $\mathcal{O}(p^3)$ in heavy baryon SU(3)
  chiral perturbation theory}.
\newblock {\em Phys. Rev. D}, 101:056021, 2020.

\bibitem{huan20202}
B.-L. Huang.
\newblock {Pion-nucleon scattering to order $p^4$ in SU(3) heavy baryon chiral
  perturbation theory}.
\newblock {\em Phys. Rev. D}, 102:116001, 2020.

\bibitem{gass1985}
J.~Gasser and H.~Leutwyler.
\newblock {Chiral perturbation theory: expansions in the mass of the strange
  qurak}.
\newblock {\em Nucl. Phys. B}, 250:465--516, 1985.

\bibitem{mojz1998}
M.~Mojzis.
\newblock {Elastic $\pi N$ scattering to $\mathcal{O}(p^3)$ in heavy baryon
  chiral perturbation theory}.
\newblock {\em Eur. Phys. J. C}, 2:181--195, 1998.

\bibitem{fett1998}
N.~Fettes, U.-G. Mei{\ss}ner, and S.~Steininger.
\newblock {Pion-nucleon scattering in chiral perturbation theory (I): Isospin
  symmetric case}.
\newblock {\em Nucl. Phys. A}, 640:199--234, 1998.

\bibitem{bora1997}
B.~Borasoy and U.-G. Mei{\ss}ner.
\newblock {Chiral expansion of baryon masses and sigma-terms}.
\newblock {\em Annals Phys.}, 254:192--232, 1997.

\bibitem{bora1999}
B.~Borasoy.
\newblock {Baryon axial currents}.
\newblock {\em Phys. Rev. D}, 59:054021, 1999.

\bibitem{olle2006}
J.~A. Oller, M.~Verbeni, and J.~Prades.
\newblock {Meson-baryon effective chiral lagrangians to $\mathcal{O}(q^3)$}.
\newblock {\em JHEP}, 09:079, 2006.
\newblock [Erratum: arXiv:hep-ph/0701096].

\bibitem{frin2006}
M.~Frink and U.-G. Mei{\ss}ner.
\newblock {On the chiral effective meson-baryon Lagrangian at third order}.
\newblock {\em Eur. Phys. J. A}, 29:255--260, 2006.

\bibitem{mach2011}
R.~Machleidt and D.~R. Entem.
\newblock {Chiral effective field theory and nuclear forces}.
\newblock {\em Phys. Rept.}, 503:1--75, 2011.

\bibitem{kais19971}
N.~Kaiser, R.~Brockmann, and W.~Weise.
\newblock {Peripheral nucleon-nucleon phase shifts and chiral symmetry}.
\newblock {\em Nucl. Phys. A}, 625:758--788, 1997.

\bibitem{erke1971}
K.~Erkelenz, R.~Alzetta, and K.~Holinde.
\newblock {Momentum space calculations and helicity formalism in nuclear
  physics}.
\newblock {\em Nucl. Phys. A}, 176:413--432, 1971.

\bibitem{pdg2018}
M.~Tanabashi $et$ $al.$.
\newblock {Review of Particle Physics}.
\newblock {\em Phys. Rev. D}, 98(10):030001, 2018.

\bibitem{chan2018}
C.~C.~Chang $et$ $al.$.
\newblock {A per-cent-level determination of the nucleon axial coupling from
  quantum chromodynamics}.
\newblock {\em Nature}, 558:91--94, 2018.

\bibitem{mark2019}
B.~M{\"a}rkisch $et$ $al.$.
\newblock {Measurement of the weak axial-vector coupling constant in the decay
  of free neutrons using a pulsed cold neutron beam}.
\newblock {\em Phys. Rev. Lett.}, 122:242501, 2019.

\bibitem{SAID}
{W. J. Briscoe $et$ $al.$, SAID on-line program, see
  http://gwdac.phys.gwu.edu}.

\bibitem{nava20131}
R.~Navarro~P\'erez, J.~E. Amaro, and E.~Ruiz~Arriola.
\newblock Partial-wave analysis of nucleon-nucleon scattering below the
  pion-production threshold.
\newblock {\em Phys. Rev. C}, 88:024002, 2013.
\newblock [Erratum Phys. Rev. C 88, 069902 (2013)].

\bibitem{nava20132}
R.~Navarro P\'erez, J.~E. Amaro, and E.~Ruiz Arriola.
\newblock Coarse-grained potential analysis of neutron-proton and proton-proton
  scattering below the pion production threshold.
\newblock {\em Phys. Rev. C}, 88:064002, 2013.
\newblock [Erratum Phys. Rev. C 91, 029901 (2015)].

\bibitem{nava20201}
R.~Navarro P\'erez and E.~Ruiz Arriola.
\newblock Uncertainty quantification and falsification of chiral nulcear
  potentials.
\newblock {\em Eur. Phys. J. A}, 56:99, 2020.

\bibitem{nava20202}
E.~Ruiz Arriola, J.~Enrique Amaro, and R.~Navarro P\'erez.
\newblock {NN} scattering and nuclear uncertainties.
\newblock {\em Front. Phys.}, 8:1, 2020.

\bibitem{simo2018}
I.~Ruiz Simo, J.~E. Amaro, E.~Ruiz Arriola, and R.~Navarro P{\'{e}}rez.
\newblock Low energy peripheral scaling in nucleon{\textendash}nucleon
  scattering and uncertainty quantification.
\newblock {\em J. Phys. G: Nucl. Part. Phys.}, 45:035107, 2018.

\bibitem{stok1993}
V.~G.~J. Stoks, R.~A.~M. Klomp, M.~C.~M. Rentmeester, and J.~J. de~Swart.
\newblock Partial-wave analysis of all nucleon-nucleon scattering data below
  350 mev.
\newblock {\em Phys. Rev. C}, 48:792--815, 1993.

\bibitem{stok1994}
V.~G.~J. Stoks, R.~A.~M. Klomp, C.~P.~F. Terheggen, and J.~J. de~Swart.
\newblock Construction of high quality nn potential models.
\newblock {\em Phys. Rev. C}, 49:2950--2962, 1994.

\bibitem{wiri1995}
R.~B. Wiringa, V.~G.~J. Stoks, and R.~Schiavilla.
\newblock Accurate nucleon-nucleon potential with charge-independence breaking.
\newblock {\em Phys. Rev. C}, 51:38--51, 1995.

\bibitem{mach2001}
R.~Machleidt.
\newblock High-precision, charge-dependent bonn nucleon-nucleon potential.
\newblock {\em Phys. Rev. C}, 63:024001, 2001.

\bibitem{gros2008}
Franz Gross and Alfred Stadler.
\newblock Covariant spectator theory of $\mathit{np}$ scattering: Phase shifts
  obtained from precision fits to data below 350 {MeV}.
\newblock {\em Phys. Rev. C}, 78:014005, 2008.

\bibitem{nava20141}
R.~Navarro~P\'erez, J.~E. Amaro, and E.~Ruiz Arriola.
\newblock Coarse-grained {NN} potential with chiral two-pion exchange.
\newblock {\em Phys. Rev. C}, 89:024004, 2014.

\bibitem{nava20142}
R.~Navarro~P\'erez, J.~E. Amaro, and E.~Ruiz Arriola.
\newblock Partial wave analysis of chiral nn interations.
\newblock {\em Few Body Syst.}, 55:983, 2014.

\bibitem{nava20143}
R.~Navarro~P\'erez, J.~E. Amaro, and E.~Ruiz~Arriola.
\newblock Statistical error analysis for phenomenological nucleon-nucleon
  potentials.
\newblock {\em Phys. Rev. C}, 89:064006, 2014.

\bibitem{nava2016}
R.~Navarro P{\'{e}}rez, J.~E. Amaro, and E.~Ruiz Arriola.
\newblock The low energy structure of the nucleon-nucleon interations:
  statistical versus systematic uncertainties.
\newblock {\em J. Phys. G: Nucl. Part. Phys.}, 43:114001, 2016.

\end{thebibliography}

\end{document}